\RequirePackage{fix-cm}
\documentclass[twocolumn,epjc3,final]{svjour3}  
\smartqed  
 \usepackage{mathrsfs}
\usepackage{amsmath, txfonts}
\usepackage{graphicx,bm,booktabs,bm}
\usepackage{longtable,lscape}
\usepackage{ulem} 
\usepackage{overpic,multirow,color}
\definecolor{cover}{rgb}{0.77,0.87,0.88}
\definecolor{blueone}{rgb}{0.1,0.1,.7}
\definecolor{citec}{rgb}{0.14,0.47,0.09}
\definecolor{two}{rgb}{0.0,0.5,0.}
\definecolor{three}{rgb}{.5,.1,0.15}
\usepackage[bookmarks=true,bookmarksopen=false,plainpages=false,breaklinks=true,
   bookmarksnumbered=true,hypertexnames=false,
   filecolor=blue,urlcolor=three,menucolor=three,
   linkcolor=three,citecolor=blueone, colorlinks,
   anchorcolor=blue,runcolor=pink,frenchlinks=red
   pdfstartview=FitH,pdftitle=title,%
   pdfauthor=author]{hyperref}
              \allowdisplaybreaks[4]

\graphicspath{{Figures/}} %

\allowdisplaybreaks[4]

\journalname{Eur. Phys. J. C}
\begin{document}
\title{Roles of  $a_0(980)^+ f_0(500,980)$ and $a_1(1260)^+\eta$ production mechanisms in  decay $D_s^+ \to \pi^+ \pi^0 \pi^0 \eta$}
\author{Jun He\thanksref{e1}
}                     
\thankstext{e1}{Corresponding author: junhe@njnu.edu.cn}
\institute{School of Physics and Technology, Nanjing
Normal University, Nanjing 210097, China}
\date{Received: date / Revised version: date}
%
\maketitle

\abstract{

In this work, we theoretically investigate the decay mechanism of $D_s^+ \to \pi^+ \pi^0 \pi^0 \eta$ based on BESIII data, considering two mechanisms: the production of two dynamically generated resonances, $D_s^+ \to a_0(980)^+ f_0(500,980)$, and a process with one dynamically generated resonance, $D_s^+ \to a_1(1260)^+\eta$. In the first mechanism, the interactions $\pi^+\eta$ and $K^+\bar{K}^0$ are included to generate the $a_0(980)^+$, while coupled-channel interactions involving $\pi^+\pi^-$, $\pi^0\pi^0$, $K^+K^-$, $K^0\bar{K}^0$, $\eta\eta$, and $\pi^0\eta$ are adopted to generate the $f_0(500)$ as well as the $f_0(980)$. For the second mechanism, the $\pi\rho$ and $\bar{K}^*{K} - K^*\bar{K}$ channels are included to generate the $a_1(1260)$. With these interactions, the amplitudes are calculated in the chiral unitary approach and used to compute the invariant mass spectra of $D_s^+ \to \pi^+ \pi^0 \pi^0 \eta$. The results suggest that the peak near 1000~\text{MeV} in the $\pi^+\eta$ invariant mass spectrum can be well interpreted as the $a_0(980)^+$ arising from a cusp effect. The $f_0(980)$, generated together with the $f_0(500)$, is responsible for the small structure near 1000~\text{MeV} in the $\pi^0\pi^0$ spectrum. The $a_1(1260)^+$ and its decay to $\rho^+\pi^0$ followed by the subsequent $\rho^+$ decay provide a good description of the $\pi^+\pi^0\pi^0$ and $\pi^+\pi^0$ mass spectra. The inclusion of an additional resonance near 800 MeV in the $\pi^+\eta$ spectrum leads to only a marginal improvement of the fit.

 } 
\section{Introduction}

The charm decay into multiple light hadrons provides a unique laboratory for studying the nature of light scalar and axial-vector mesons, whose internal structures remain controversial. The more hadrons in the final state, the richer the underlying dynamics, as more intermediate resonances and coupled-channel effects can be involved and interfere with one another. The recent high-statistics data from the BESIII collaboration on the four-body decay $D_s^+ \to \pi^+ \pi^0 \pi^0 \eta$ offer a promising platform, where the simultaneous production of two resonances (e.g., $a_0(980)$ and $f_0(500)$), alongside single-resonance processes  $a_1(1260)^+\eta$, can be explored~\cite{BESIII:2026mtz}. Such a multi-hadron final state provides valuable clues for understanding the dynamical origin of these resonances and the role of coupled-channel effects in a complex environment.

The nature of these light scalar and axial-vector mesons has long been a subject of debate~\cite{Pelaez:2015qba,Gross:2022hyw,Close:2002zu,Amsler:2004ps}. 
While a simple $q\bar q$ assignment fails to account for many of their observed properties, their masses, decay widths, and coupling patterns suggest more complex configurations, such as tetraquarks, or dynamically generated resonances from coupled-channel interactions~\cite{Oller:1997ti,Roca:2005nm,Jaffe:1975fd,Tornqvist:1995kr,Branz:2007xp,Clymton:2022jmv,Weinstein:1982gc,Nagahiro:2011jn,Achasov:1980tb,Baru:2003qq}. 
In particular, it is noteworthy that the $a_0(980)$/$f_0(980)$ and the $a_1(1260)$ are located close to the corresponding strange-meson pair thresholds: $K\bar K$ for the scalars and $K\bar K^*$ for the axial-vector. This proximity strongly suggests that these states may originate from these channels with possible coupled-channel dynamics.

Indeed, the $a_0(980)$ is widely interpreted as a $K\bar K$ bound state or as a cusp effect triggered by the $K\bar K$ threshold, while the $f_0(980)$ shares a similar molecular interpretation~\cite{Oller:1997ti,Branz:2007xp,Weinstein:1990gu}. Both resonances emerge naturally from the coupled-channel $S$-wave scattering involving $K\bar K$ and other pseudoscalar channels such as $\pi\eta$  and/or  $\pi\pi$~\cite{Oller:1997ti,Ahmed:2020kmp,Dudek:2016cru}. In contrast, the broad $f_0(500)$ resonance arises primarily from the strong $\pi\pi$ interactions in the isoscalar-scalar channel, with other coupled channels taken into account~\cite{Oller:1997ti,Danilkin:2020pak,Pelaez:2025jrn}. On the other hand, the axial-vector $a_1(1260)$ exhibits an analogous dynamical origin: its mass lies close to the $K\bar K^*$ threshold, and it couples strongly to both the $\pi\rho$ and $K\bar K^*$ channels. It is therefore naturally interpreted as a dynamically generated resonance from the coupled-channel interactions involving a pseudoscalar and a vector meson~\cite{Roca:2005nm,Clymton:2022jmv,Janssen:1994uf}.

These light scalar and axial-vector mesons have been extensively studied in charmed meson decays, particularly in three-body final states~\cite{Oset:2016lyh,Lin:2021isc,Duan:2024czu,Lyu:2025oow}. There are also a few studies on four-body decays, for example, the decay $D_s^+ \to \pi^+ \pi^+ \pi^- \eta$~\cite{Song:2022kac}, which is quite similar to the process considered in the present work. However, the four-body decay $D_s^+ \to \pi^+ \pi^0 \pi^0 \eta$ offers a valuable opportunity to investigate their interplay and coupled-channel dynamics. Two distinct mechanisms are suggested by the experimental analysis: (i) the production of two dynamically generated resonances, $D_s^+ \to a_0(980)^+ f_0(500)$ (with the $f_0(980)$ also considered in the present work), and (ii) the production of a single dynamically generated resonance, $D_s^+ \to a_1(1260)^+ \eta$, followed by the cascade decay $a_1(1260)^+ \to \rho^+ \pi^0 \to \pi^+ \pi^0 \pi^0$. The relevant coupled-channel interactions have been well described within the chiral unitary approach~\cite{Oller:1997ti,Roca:2005nm,Lin:2021isc,Duan:2024czu}, which makes it convenient to use this framework to analyze the BESIII data for the decay $D_s^+ \to \pi^+ \pi^0 \pi^0 \eta$.

The paper is organized as follows. In Sec.~\ref{mechnism}, we present the chiral unitary formalism for the two mechanisms and define the amplitudes for the decay. In Sec.~\ref{results}, we show our numerical results for the invariant mass spectra and compare them with the BESIII data. Finally, a summary and discussion are given in Sec.~\ref{summary}.

\section{Decay mechanism}\label{mechnism}

\begin{figure}[h!]
  \includegraphics[bb=80 630 540 770,clip,scale=0.52]{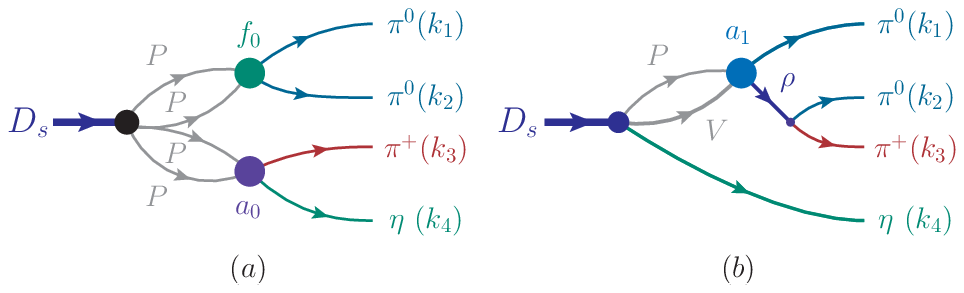}
  \caption{Feynman diagrams for $D_s^+ \to \pi^+ \pi^0 \pi^0\eta$. Panel (a) shows the rescattering mechanism involving two rescatterings that produce the $a_0(980)^+$ and $f_0(500,980)$. Panel (b) illustrates the rescattering mechanism with a single rescattering for the $a_1(1260)$.}
  \label{diagram}
\end{figure}

The BESIII data exhibit a prominent peak in the $\pi^+\eta$ invariant mass spectrum, as well as a peak in the $\pi^+\pi^0$ spectrum. The experimental analysis suggests that these structures originate from the $a_0(980)^+$ accompanied by the $f_0(500)$ and the $\rho^+$ meson, with the latter arising from an intermediate $a_1(1260)^+$ state. In the present work, we also adopt this phenomenological picture, but treat these resonances as dynamically generated states from the final-state rescatterings, as depicted in Fig.~\ref{diagram}.

The primary mechanism emphasized by the BESIII collaboration is the production of two resonances simultaneously, i.e., $D_s^+ \to a_0(980)^+ f_0(500)$. This mechanism is responsible for the sharp peak in the $\pi^+\eta$ spectrum and the broad enhancement in the $\pi^0\pi^0$ spectrum. In our approach, we generate the $a_0(980)^+$ dynamically from the coupled-channel interactions of $(PP)^+=\pi^+\eta$ and $K^+\bar{K}^0$ (hereafter denoted as the $a_0$ system for simplicity), while the $f_0(500)$ is obtained from the coupled channels $(PP)^0=\pi^+\pi^-$, $\pi^0\pi^0$, $K^+K^-$, $K^0\bar{K}^0$, $\eta\eta$, and $\pi^0\eta$ (denoted as the $f_0$ system for simplicity). Concretely, the mechanism proceeds as follows: the $D_s^+$ decays first into four pseudoscalar mesons; among them, two mesons with positive charge $(PP)^+$ rescatter to form the $a_0(980)^+$, which subsequently decays into $\pi^+\eta$, while the remaining two mesons with neutral charge $(PP)^0$ rescatter to form the $f_0(500)$, which then decays into $\pi^0\pi^0$. It is worth noting that the coupled-channel interactions for the $f_0$ system may also generate a state or a cusp near 1000~MeV in the $\pi^0\pi^0$ spectrum, often identified as the $f_0(980)$, originating from the $K^+K^-$ and $K^0\bar{K}^0$ channels. This contribution may correspond to the small structure near 1000~MeV observed in the experimental $\pi^0\pi^0$ spectrum and partially accounts for the $D_s^+ \to \pi^+ (\pi^0 \pi^0)_{\rm S-wave}\,\eta$ process considered in the experimental analysis.

Since all mesons involved are spin-zero, the decay amplitude can be easily written as
\begin{align}
{\cal M}_{a_0f_0}&=\sum_{(PP)^+}\sum_{(PP)^0}\int \frac{d^4k}{(2\pi)^4} \frac{d^4k'}{(2\pi)^4} T_{(PP)^+\to \pi^+\eta}(k) G(k) \nonumber \\ 
&\cdot T_{(PP)^0\to \pi^0\pi^0 }(k')   G(k') A_{(PP)^+(PP)^0},
\end{align}
where $k$ and $k'$ are the loop momenta. In the present work, we adopt the chiral unitary approach~\cite{Oller:2000ma,Oset:2016lyh}, in which the rescattering amplitudes $T$ depend only on the invariant masses $s$ and $s'$ of the corresponding two-meson subsystems, and consequently can be factored out of the integrations.The amplitude then reduces to
\begin{align}
{\cal M}_{a_0f_0}&= \sum_{(PP)^+}T_{(PP)^+\to \pi^+\eta}(s) G(s)  \nonumber \\ 
&\cdot\sum_{(PP)^0}T_{(PP)^0\to \pi^0\pi^0 }(s')  G(s') A_{(PP)^+(PP)^0},
\end{align}
where $A_{(PP)^+(PP)^0}=g_{(PP)^+(PP)^0}$ is a constant vertex factor representing the primary decay $D_s^+ \to (PP)^+ (PP)^0$.

The $G$ function in the cut-off regularization is given by~\cite{Oller:1997ti}
\begin{align}
G(\sqrt{s}) = \int_0^{q_{\text{max}}} \frac{q^2\,dq}{(2\pi)^2}
\frac{\omega_1 + \omega_2}{\omega_1 \omega_2 \left[ s - (\omega_1 + \omega_2)^2 + i\epsilon \right]},
\label{eq:G}
\end{align}
with $\omega_i = \sqrt{m_i^2 + q^2}$ and $\sqrt{s}$ the center-of-mass energy of the two mesons in the loop, and $q_{\text{max}}$ stands for the maximum value of the modulus of the three-momentum $q$ allowed in the integral of Eq.~\eqref{eq:G}. The rescattering amplitudes $T$ are obtained by solving the coupled-channel Bethe–Salpeter equation in the on-shell factorized form,
\begin{align}
T= [1-VG]^{-1}V,
\label{eq:T}
\end{align}
where ${V}$ denotes the matrix of the potential kernel. The explicit forms of ${V}$ for both the positively charged and neutral channels have been provided and are widely used in the literature~\cite{Oller:1997ti,Lin:2021isc,Duan:2024czu}; hence we do not repeat them here.

The second mechanism is the decay $D_s^+ \to a_1(1260)^+\,\eta$, where the $a_1(1260)^+$ is dynamically generated from the interaction with channels $(PV)^+=(\pi\rho)^+$ and $\frac{1}{\sqrt{2}}(\bar{K}^*{K} - K^*\bar{K})^+$  (denoted as the $a_1 $ system for simplicity) in our framework. After rescattering, the intermediate $\rho^+$ meson decays into $\pi^0\pi^+$. In this case, we do not attempt to describe the internal structure of the $\rho^+$ meson; instead, we parametrize its lineshape directly using the Gounaris–Sakurai formula as done in the experimental analysis~\cite{BESIII:2026mtz}.

The decay amplitude can be written as 
\begin{align}
{\cal M}_{a_1}&=\Gamma_{\rho^+\to\pi^+\pi^0}G_{\rho^+} (s_{\rho^+})\sum_{(PV)^+}\int \frac{d^4k}{(2\pi)^4} T_{(PV)^+\to \pi^0\rho^+}(k) \nonumber \\ 
&\cdot G_{(PV)^+}(k) A_{D_s^+\to (PV)^+\eta},
\end{align}
where $G_{\rho^+}(s_{\rho^+})$ is the Gounaris--Sakurai propagator with $s_{\rho^+}$ being the invariant mass squared of the $\rho^+$ meson. The decay vertex of the $\rho^+$ can be written as
\begin{align}
\Gamma_{\rho^+\to\pi^+\pi^0}
= g_\rho ({\bm k}_{\pi^+} - {\bm k}_{\pi^0}) \cdot {\bm \epsilon}_{\rho^+},
\end{align}
where $k_{\pi^+}$ and $k_{\pi^0}$ are the momenta of the final $\pi^+$ and $\pi^0$ mesons, respectively, and $\epsilon_{\rho^+}$ is the polarization vector of the $\rho^+$.

The vertex for the $D_s^+ \to (VP)^+ \eta$ decay is expressed as
\begin{align}
A_{D_s^+\to (VP)^+\eta}=g_{PV\eta}({\bm k}_{\eta}-{\bm k}_{(PV)^+})\cdot{\bm \epsilon}_V, 
\end{align}
where $k_{\eta}$ and $k_{(PV)^+}$ are the momenta of the final $\eta$ and the rescattering $(PV)^+$ system, respectively, and $\epsilon_V$ denotes the polarization vector of the vector meson.

In the chiral unitary approach, the scattering matrix for transverse polarization modes of the external vector mesons leads to the following unitarized amplitude~\cite{Roca:2005nm}:
\begin{equation}
{T} = \left[ {1} +{V} \hat{{G}} \right]^{-1} (-{V})~{\bm \epsilon}\cdot {\bm \epsilon}',
\label{eq:T_trans}
\end{equation}
where $\hat{{G}}$ is a diagonal matrix with its $l$-th element given by
$\hat{G}_{l} = G_l \left( 1 - \frac{1}{3} \frac{q^2}{M_l^2} \right)$
and $G_l$ is the two-meson loop function for the $l$-th channel containing a vector and a pseudoscalar meson, which has the same form as Eq.~(\ref{eq:G}). The explicit form of the potential is also provided in Ref.~\cite{Roca:2005nm}.

Thus, the decay amplitude for the $a_1$ mechanism is obtained as
\begin{align}
{\cal M}_{a_1}&= g_\rho ({\bm k}_{\pi^+}-{\bm k}_{\pi^0})\cdot ({\bm k}_{\eta}-{\bm k}_{(PV)^+}) G_\rho (s_\rho) \nonumber\\
&\cdot \sum_{(PV)^+} g_{PV\eta} T_{a_1}(s) G_{(VP)^+}(s),
\end{align}
with symmetrization implemented by exchanging the two $\pi^0$ mesons.

\begin{figure}[h!]
  \includegraphics[bb=80 390 540 770,clip,scale=0.5]{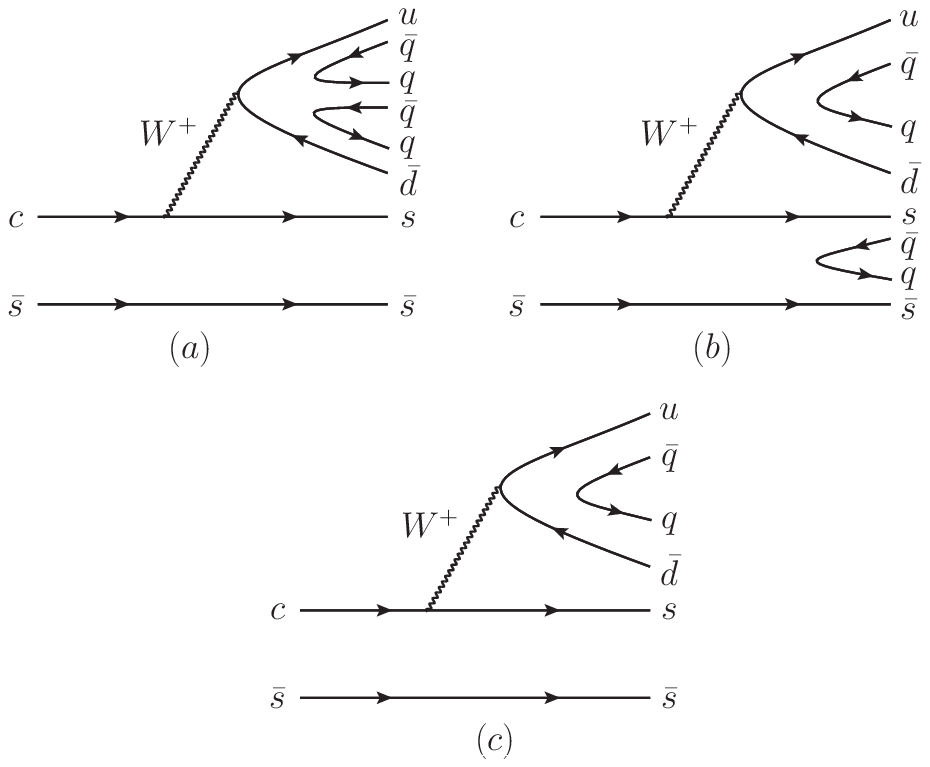}
  \caption{Quark-level diagrams for the $D_s^+$ decay. Panels (a) and (b) correspond to the $D_s^+ \to (PP)^+ (PP)^0$ processes, while panel (c) corresponds to $D_s^+ \to (PV)\,\eta$.}
  \label{decay}
\end{figure}

We now turn to the determination of the effective coupling constants $g_{(PP)^+(PP)^0}$ and $g_{PV\eta}$, which encode the primary weak decay strengths of the $D_s^+$ into two-meson pairs. Following the procedure adopted in Ref.~\cite{Oset:2016lyh}, their relative ratios among channels within the same sector are constrained by SU(3) flavor symmetry, as shown by the quark-level diagrams for the $D_s^+$ decay presented in Fig.~\ref{decay}. The $D_s^+$ is a $c\bar{s}$ bound state, and its weak decay proceeds through the $c \to s u \bar{d}$ transition. At the quark level, the $s\bar{s}$ pair from the initial $\bar{s}$ and the $u\bar{d}$ pair from the weak vertex hadronize into final mesons. Diagrams (a) and (b) correspond to the $(PP)^+(PP)^0$ mechanism. Diagram (c) corresponds to the $PV\eta$ mechanism, with $(PV)^+$ channels. Moreover, we note that the internal emission diagrams are suppressed relative to the external emission mechanism; for simplicity, we retain only the latter in the present work.

To implement the $SU(3)$ flavor symmetry in the hadronization, we adopt the standard pseudoscalar and vector meson matrices. Following the mixing scheme of Ref.~\cite{Bramon:1992kr}, the $P$ matrix for the pseudoscalar nonet is written as
\begin{align}
P =
\begin{pmatrix}
\frac{\pi^0}{\sqrt{2}} + \frac{\eta}{\sqrt{3}} + \frac{\eta'}{\sqrt{6}}
& \pi^+ & K^+ \\[8pt]
\pi^- & -\frac{\pi^0 }{\sqrt{2}}+ \frac{\eta}{\sqrt{3}} + \frac{\eta'}{\sqrt{6}} & K^0 \\[8pt]
K^- & \bar{K}^0 & -\frac{\eta}{\sqrt{3}} + \frac{2\eta'}{\sqrt{6}}\,
\end{pmatrix}.
\label{eq:Pmatrix}
\end{align}
When the $\eta$--$\eta'$ mixing is taken into account, the physical $\eta$ and $\eta'$ fields are mixtures of the octet and singlet states. In practice, the contributions from the $\eta'$ are not directly involved in the current work; nevertheless, we include them consistently in the calculation for completeness. The vector-meson matrix $V$ is defined as
\begin{equation}
V =
\begin{pmatrix}
\dfrac{1}{\sqrt{2}}\rho^0 + \dfrac{1}{\sqrt{2}}\omega
& \rho^+ & K^{*+} \\[8pt]
\rho^- & -\dfrac{1}{\sqrt{2}}\rho^0 + \dfrac{1}{\sqrt{2}}\omega & K^{*0} \\[8pt]
K^{*-} & \bar{K}^{*0} & \phi
\end{pmatrix},
\label{eq:Vmatrix}
\end{equation}
which will be used in the construction of the $PV$ production amplitudes.

For the mechanism in Fig.~\ref{decay} (a), the $u\bar{d}$ pair hadronizes into three pseudoscalar mesons, while the $s\bar{s}$ pair forms an $\eta$ meson. Therefore, the amplitude for this process can be written as
\begin{align}
D_s\to(s\bar{s})(u\bar{d})& \to(s\bar{s})\,(u q\bar{q} q\bar{q} \bar{d})
 \to g_a~\eta\,(PPP)_{12} \nonumber \\
= g_a\Bigg[& \left(\frac{1}{2}\pi^0\pi^0 + \pi^+\pi^- + \eta\eta + K^+K^- + K^0\bar{K}^0 \right)\pi^+\eta
\nonumber\\
&+ \frac{1}{\sqrt{3}}\, \eta\eta \, K^+\bar{K}^0 \Bigg],
\label{eq:PP_amp_noeta}
\end{align}
where the subscript 12 of $(PPP)_{12}$ indicates the combination that produces a positively charged meson pair $(PP)^+$ from the $u\bar{d}$ quark pair. In Eq.~\eqref{eq:PP_amp_noeta}, the first term corresponds to $(PP)^+ = \pi^+\eta$ and $(PP)^0$ being any of the neutral scalar combinations ($\pi^0\pi^0$, $\pi^+\pi^-$, $\eta\eta$, $K^+K^-$, $K^0\bar{K}^0$), each with the indicated relative weight. The second term corresponds to $(PP)^+ = K^+\bar{K}^0$ and $(PP)^0 = \eta\eta$, with a relative factor $1/\sqrt{3}$. The overall constant $g_a$ is the same for all these channels, reflecting the common production mechanism in Fig.~\ref{decay} (a).

For the mechanism in Fig.~\ref{decay} (b), the two quark pairs are again the $u\bar{d}$ from the $W$ boson and the $s\bar{s}$ from the initial $\bar{s}$. Here, the $s\bar{s}$ pair hadronizes into a neutral two-meson pair $(PP)^0_{33}$, while the $u\bar{d}$ pair hadronizes into a positively charged pair $(PP)^+_{12}$. The amplitude is given by
\begin{align}
D_s&\to(s\bar{s})(u\bar{d}) \to g_b ~(PP)^0_{33}\,(PP)^+_{12}\nonumber\\
&= g_b \Big( K^-K^+ + \bar{K}^0 K^0 + \frac{1}{3}\eta\eta \Big)
\Big( \frac{2}{\sqrt{3}}\,\eta\pi^+ + K^+\bar{K}^0 \Big),
\label{eq:PP33_12}
\end{align}
where $(PP)^0_{33}$ denotes a neutral pair formed from the $s\bar{s}$ component (e.g., $K^-K^+$, $\bar{K}^0 K^0$, or $\eta\eta$), and $(PP)^+_{12}$ is again the positively charged pair from the $u\bar{d}$ component. The constant $g_b$ is another effective coupling that, in principle, could differ from $g_a$ due to different color-flow topologies. The expression in Eq.~\eqref{eq:PP33_12} explicitly shows the allowed combinations: for $(PP)^+$ we have $\eta\pi^+$ and $K^+\bar{K}^0$ with weights $2/\sqrt{3}$ and $1$, respectively; for $(PP)^0$ we have $K^-K^+$, $\bar{K}^0 K^0$, and $\eta\eta$ with weights $1$,$1$, and $1/3$, respectively.

There is, naturally, a third quark-level diagram, where both additional $q\bar{q}$ pairs are created from the $s\bar{s}$ system (rather than from the $u\bar{d}$ or split between the two). In this topology, the $u\bar{d}$ pair from the $W$ boson directly forms a $\pi^+$ meson, while the $s\bar{s}$ system, after creating two extra $q\bar{q}$ pairs, hadronizes into a neutral pair and an $\eta$ meson, resulting in the combination $(K^-K^+ + \bar{K}^0 K^0 + \frac{1}{3}\eta\eta)\,\eta\pi^+$. This mechanism therefore yields only the $(PP)^+ = \eta\pi^+$ term, without the $K^+\bar{K}^0$ component that appears in the mechanism of Fig.~\ref{decay} (b). If included, the effect of this diagram is to modify the relative weight of the $\frac{2}{\sqrt{3}}\eta\pi^+$ and $K^+\bar{K}^0$ terms in Eq.~\eqref{eq:PP33_12}, thereby introducing an extra free parameter. However, the creation of extra $q\bar{q}$ pairs from the $s\bar{s}$ system is dynamically suppressed compared with the analogous process from the $u\bar{d}$ pair. This is because the $u\bar{d}$ pair originates from the decay of the heavy $W$ boson and thus carries sufficient energy to readily produce additional $q\bar{q}$ pairs during hadronization, whereas the $s\bar{s}$ system lacks such an energetic source and is therefore less favorable for producing extra quark pairs. Given the smallness of this effect, we neglect it in the present work.

Finally, for the $D_s^+ \to (PV)^+\eta$ mechanism, we analogously write the production amplitude as a sum over the relevant $(PV)^+$ channels, with a coupling $g_{PV\eta}$ that again serves as a free parameter. Since only one quark pair is required, this mechanism is simpler than the above four-pseudoscalar decay processes, and can be written as
\begin{align}
D_s \to (s\bar{s})(u\bar{d}) &\to (s\bar{s})(u q\bar{q} \bar{d}) =\frac{g_c}{\sqrt{2}} \eta\,[(PV)^+,(VP)^+]_{12} \nonumber \\
= \frac{g_c}{\sqrt{2}}\,\eta\Big[ &\frac{1}{\sqrt{2}}(\pi^0 \rho^+ - \pi^+ \rho^0) + K^+ \bar K^{*0}, \nonumber \\
&\frac{1}{\sqrt{2}}(\rho^0 \pi^+ - \rho^+ \pi^0) + K^{*+} \bar K^0 \Big] ,
\end{align}
where the notation $[(PV)^+,(VP)^+]_{12}$ denotes the combination of a pseudoscalar and a vector meson that carries charge $+1$ from the $u\bar{d}$ pair. In the above expression, the two lines inside the brackets correspond to the two possible orderings, which are both included to account for the vector–pseudoscalar mixing. For the $a_1(1260)^+$ resonance considered in the present work, we adopt the specific combination that yields the correct $G$-parity~\cite{Dai:2018thd}:
\begin{equation}
H = g_c\,\eta\Big[ \sqrt{2}\frac{1}{\sqrt{2}}(\pi^0 \rho^+ - \pi^+ \rho^0) + \frac{1}{\sqrt{2}}(K^+ \bar K^{*0} - K^{*+} \bar K^0) \Big] .
\end{equation}
Furthermore, a quark-level diagram where the extra $q\bar{q}$ pair emerges from the $s\bar{s}$ system is also possible in principle, but the resulting processes are not relevant to the decay topology considered in the current work. Therefore, we do not take them into account.

Hence, besides the relative ratios, the effective couplings $g_{(PP)^+(PP)^0}$ and $g_{PV\eta}$ are replaced by $g_a$, $g_b$, and $g_{a_1}$ for the three distinct mechanisms, respectively. Specifically, we extract the coupling constants from the amplitudes $\mathcal{M}$ so that the total amplitude for the above mechanisms can be written as
\begin{align}
\mathcal{M} &= g_a e^{i\phi_a} \mathcal{M}_{a_0f_0}^{a}
          + g_b e^{i\phi_b} \mathcal{M}_{a_0f_0}^{b}
          + g_{a_1} e^{i\phi_{a_1}} \mathcal{M}_{a_1}
          + \mathcal{M}_{\rm bk},
\label{eq:total_amplitude}
\end{align}
where $e^{i\phi}$ denotes the relative phase for each intermediate mechanism, and $\mathcal{M}$ represents the corresponding reduced amplitude (with the global coupling factor removed). We define $g_{a_1}$ by absorbing the $\rho$ decay constant $g_\rho$ into $g_c$, so the latter is not explicitly separated in our parametrization. The first two terms correspond to the two different quark-level topologies for the $D_s^+ \to (PP)^+(PP)^0$ rescattering mechanism, as depicted in panels (a) and (b) of Fig.~\ref{decay}, respectively. The third term accounts for the $D_s^+ \to (PV)^+\eta$ mechanism (panel (c) of Fig.~\ref{decay}). The background contribution $\mathcal{M}_{\rm bk}$ is treated as a constant $g_{\rm bk}$, which effectively absorbs all nonresonant contributions.

In the experimental data of the $\pi^+\eta$ invariant mass spectrum, there exists a peak near 800~MeV that was not included in the fitting of the experimental analysis. In our present theoretical approach, there is also no dynamical mechanism to account for this structure. Hence, we simulate it by introducing a Breit–Wigner resonance $X(800)$ of the form
\begin{align}
\mathcal{M}_X = \frac{g_X \, e^{i\phi_X}}{s_{\pi^+\eta} - M_X^2 + i M_X \Gamma_X},
\label{eq:BW}
\end{align}
where $g_X$ and $e^{i\phi_X}$ denote the strength and phase, while $M_X$ and $\Gamma_X$ are the mass and width, respectively. All these parameters are treated as free in the fit. In our calculations, we will consider both scenarios with and without this additional resonance, in order to assess its possible impact on the description of the data.

\section{Numerical Result}\label{results}

\subsection{Fittting Procedure}

With the decay amplitude at hand, the differential decay width is expressed as
\begin{equation}
d\Gamma = \frac{1}{2M_{D_s}} \sum |\mathcal{M}|^2 \, d\Phi,
\end{equation}
where $M_{D_s}$ denotes the mass of the initial $D_s$ meson, and $d\Phi$ is the Lorentz-invariant phase-space element. Following our previous work~\cite{He:2025vij,He:2026vlf}, the phase-space integration is carried out using the Monte Carlo method proposed in Ref.~\cite{James:1968gu}, which we implement in the Julia programming language. The corresponding package, \texttt{DalitzPlot.jl}, is publicly available on GitHub~\cite{code} and provides a simulation framework for multi-body decays. A total of $10^7$ Monte Carlo events are generated, with final-state momenta sampled under strict enforcement of energy–momentum conservation. From these simulated events, the invariant mass sprectra are constructed and subsequently used in the analysis.

We will fit the BESIII invariant mass spectra of $\pi^+\eta$, $\pi^0\pi^0$, $\pi^+\pi^0\pi^0$, $\eta\pi^0$, $\pi^+\pi^0$, $\eta\pi^0\pi^+$, and $\eta\pi^0\pi^0$. As summarized in Table~\ref{tab:fit}, in the fitting, we take the strength $g$ and phase $\phi$ for the three decay mechanisms shown in Fig.~\ref{decay} as free parameters. The background is described by a real constant $g_{\rm bk}$. These seven parameters are used for all fits.
In the present work, we consider three interaction systems $a_0$, $f_0$, and $a_1$ (We note that these three interaction systems do not directly correspond one-to-one to the three decay mechanisms in Fig.~\ref{decay}). For each interaction, we need a cutoff $q_{\rm max}$. In the literature, for the $a_0$ system, $q_{\rm max}$ is chosen as 600~MeV to reproduce the experimental data of $\Lambda_c^+ \to \eta \pi^+ \Lambda$~\cite{Duan:2024czu}. For the $f_0$ system, $q_{\rm max} = 650$~MeV is adopted to fit $D_s^+ \to a_0(980)\, e^+ \nu_e$~\cite{Lin:2021isc}. In this work, we will perform fits with these two parameters either fixed to the above values or set free (denoted as ``Fixed'' and ``Free'' in Table~\ref{tab:fit}). For the $a_1(1260)$ channel, a cutoff around 1000~MeV is  used in the literature~\cite{Roca:2005nm}. However, when fitting the BESIII data, we find that the fitted value of $q_{\rm max}$ tends to become very small; to keep it within a reasonable range, we fix $q_{\rm max} = 500$~MeV in all fits.
We also provide results with and without a resonance $X(800)$ near 800~MeV in the $\pi^+\eta$ spectrum, parametrized as a Breit–Wigner form, which introduces four parameters: $g_X$, $\phi_X$, $M_X$, and $\Gamma_X$. However, since the uncertainty of the data near 800~MeV is very large, leaving all four parameters free would allow the fit to absorb structures in other spectra rather than describing this specific peak. Therefore, we fix the width to $\Gamma_X = 50$~MeV to ensure that it properly describes the peak near 800~MeV.

\renewcommand\tabcolsep{0.22cm}
\renewcommand{\arraystretch}{1.45}
\begin{table}[h!]
\caption{Fit parameters and $\chi^2$ values for the four scenarios with fixed/free $q_{\rm max}$ and with/without $X(800)$. The parameters $g$ denote relative strengths (only magnitudes are physically meaningful), $\phi$ are relative phases (in unit of $\pi$), and $q_{\rm max}$ values are given in GeV. The masses $M_X$ and widths $\Gamma_X$ of the $X(800)$ resonance are in MeV. Fixed values are shown in italics. The last row gives the total $\chi^2$ per degree of freedom ($\chi^2_{\rm tot}/{\rm ndf}$) for each fit. \label{tab:fit}}
\begin{tabular}{ccccc}
		\toprule
Fits           &Fixed                     &Fixed(X)  &Free   & Free(X)   \\\hline
$g_a$          &$1.851_{-0.086}^{+0.092}$&$4.112_{-0.082}^{+0.079}$&$12.44_{-0.06}^{+0.62}$   &$8.249_{-0.037}^{+0.042}$    \\
$\phi_a$       &$0.713_{-0.032}^{+0.035}$&$0.868_{-0.024}^{+0.020}$&$1.374_{-0.005}^{+0.005}$ &$1.224_{-0.006}^{+0.008}$    \\
$g_b$          &$3.870_{-0.188}^{+0.175}$&$4.310_{-0.363}^{+0.262}$&$6.613_{-0.060}^{+0.061}$ &$4.614_{-0.041}^{+0.040}$    \\
$\phi_b$       &$1.356_{-0.006}^{+0.006}$&$1.512_{-0.005}^{+0.005}$&$0.344_{-0.008}^{+0.008}$ &$0.174_{-0.008}^{+0.007}$    \\
$g_{a_1}$      &$2.314_{-0.056}^{+0.063}$&$2.217_{-0.062}^{+0.056}$&$2.147_{-0.063}^{+0.055}$ &$2.326_{-0.053}^{+0.058}$    \\
$\phi_{a_1}$   &$1.546_{-0.035}^{+0.040}$&$1.522_{-0.039}^{+0.031}$&$1.792_{-0.028}^{+0.024}$ &$1.688_{-0.028}^{+0.028}$    \\
$g_{\rm bk}$   &$3.821_{-0.035}^{+0.037}$&$3.222_{-0.048}^{+0.036}$&$3.735_{-0.082}^{+0.080}$ &$2.649_{-0.071}^{+0.090}$    \\
$q^{a_0}_{\rm max}$&$\it 0.6$            &$\it 0.6$                &$0.760_{-0.006}^{+0.006}$ &$0.896_{-0.008}^{+0.009}$    \\
$q^{f_0}_{\rm max}$&$\it 0.65$           &$\it 0.65$               &$0.854_{-0.009}^{+0.009}$ &$0.913_{-0.012}^{+0.016}$    \\
$q^{a_1}_{\rm max}$&$\it 0.5$            &$\it 0.5$                &$\it 0.5$                 &$\it 0.5$  \\
$g_X$          &$--$                     &$0.108_{-0.007}^{+0.006}$&$--$                      &$0.105_{-0.004}^{+0.009}$    \\
$\phi_X$       &$--$                     &$0.220_{-0.036}^{+0.040}$&$--$                      &$0.432_{-0.072}^{+0.067}$    \\
$M_X$          &$--$                     &$801.7_{-7.4}^{+5.7}$     &$--$                     &$0.815_{-0.004}^{+0.010}$    \\
$\Gamma_X$     &$--$                     &$\it 50.0$               &$--$                      &$\it 50.0$   \\
${\chi^2_{\rm tot}}/{\rm ndf}$ &$1.447$  &$1.398$                  &$1.244$                   &$1.190$  \\
		\bottomrule
	\end{tabular}
\end{table}

\subsection{Results in scenario Fixed}

\begin{figure}[h!]
\centering
\includegraphics[bb=5 0 800 1100,clip,scale=0.367]{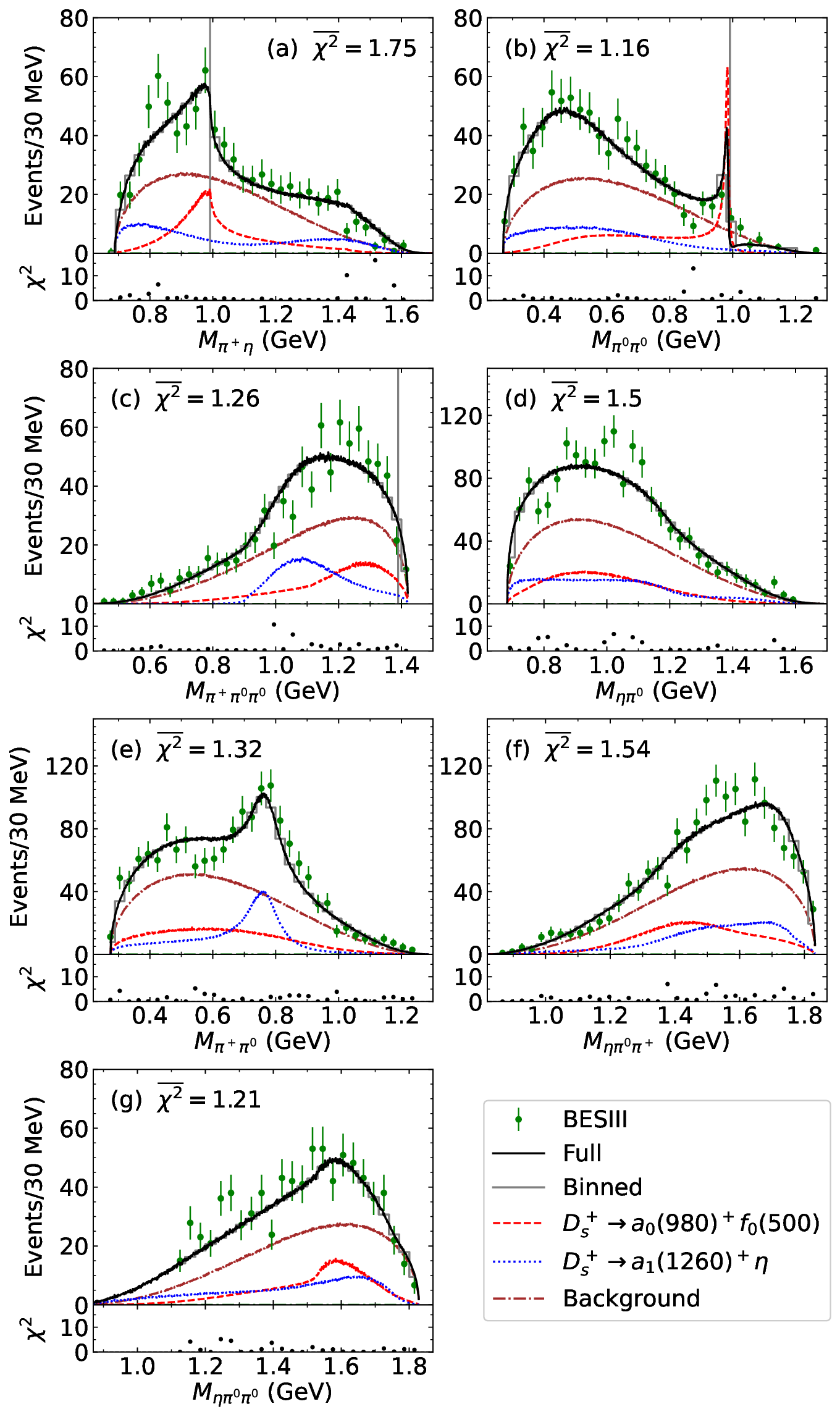}
\caption{The invariant mass spectra of the decay $D_s^+ \to \pi^+ \pi^0 \pi^0\eta$ in scenario Fixed. The solid black curve corresponds to the full model, the red dashed curve to the $D_s^+ \to a_0(980)^+ f_0(500,980)$ mechanism, the blue dotted curve to the $D_s^+ \to a_1(1260)^+ \eta$ mechanism, and the brown dash-dotted curve to the background. The binned results are shown as a gray histogram. The data points are taken from the BESIII Collaboration~\cite{BESIII:2026mtz}. The scatter points in the lower part of each panel represent the individual $\chi^2$ contributions of the corresponding data points. The horizontal dashed lines in panels (a), (b), and (c) indicate the thresholds of $K^+\bar{K}^0$, $K\bar{K}$, and $K\bar{K}^*$, respectively.}
\label{fig:Fixed}
\end{figure}

In this section, we present the results obtained with predetermined values of $q_{\rm max}$ for the three interaction systems, without including the $X(800)$ resonance, as listed in the second column of Table~\ref{tab:fit}. The $\chi^2/{\rm ndf}$ is 1.447, indicating a reasonably good fit with only seven parameters. The explicit results are illustrated in Fig.~\ref{fig:Fixed}, where the contributions from the individual mechanisms are also shown. Generally speaking, all seven invariant mass spectra are well reproduced in this model. 

The $\pi^+\eta$ spectrum involves the $a_0$ system with coupled channels $\pi^+\eta$ and $K^+\bar{K}^0$. In this two-channel system, the individual single-channel interactions play only a minor role; instead, the coupled potential between the two channels dominates the dynamics. This strong coupling leads to a significant energy dependence of the scattering amplitude near the $K^+\bar{K}^0$ threshold. In the complex plane, there exists a pole at $(1122 - 39i)$~MeV in the second Riemann sheet of the $\pi^+\eta$ channel and the first sheet of $K^+\bar{K}^0$, which arises mainly from the coupled potential. However, its direct contribution to the physical spectrum is negligible. Consequently, the observed sharp peak in the $\pi^+\eta$ spectrum cannot be attributed to a resonant pole. Instead, it is generated as a cusp effect at the $K^+\bar{K}^0$ threshold, where the opening of the $K^+\bar{K}^0$ channel produces a characteristic threshold peak in the $\pi^+\eta$ amplitude. Thus, the $a_0(980)^+$ is a manifestation of the coupled-channel dynamics in the vicinity of the threshold, rather than a genuine resonance.

In the $\pi^0\pi^0$ spectrum, the structure near 1000~MeV is attributed to the dynamically generated $f_0(980)$ state, with a pole at $(985 - 6i)$~MeV, and the broad enhancement at lower energies is due to the $f_0(500)$ state with a pole at $(458 - 244i)$~MeV; both pole positions are the same as  those in Ref.~\cite{Lin:2021isc} because we adopt the same potential and $q_{\rm max}$ values. Since the $f_0(980)$ pole lies very close to the $K\bar{K}$ threshold, the corresponding peak is extremely sharp. After binning, the results agree well with the data. The broad enhancement at lower energies arises from the $f_0(500)$ pole, which provides a broad contribution because of its large width.

For the $\pi^+\pi^0\pi^0$ spectrum, with a fixed $q_{\rm max} = 0.5$~GeV, the $a_1(1260)^+$ is dynamically generated from the $(\pi\rho)^+$ and $(\bar{K}^*{K} - K^*\bar{K})^+$  interaction, with a pole at $(1010 - 175i)$~MeV, producing a broad bump around 1050~MeV. Compared with results obtained using $q_{\rm max} = 1000$~MeV in Ref.~\cite{Roca:2005nm}, the real part of the pole position is slightly smaller, while the imaginary part is about twice as large. Since the experimental bump is located around 1.2~GeV, the fit tends to push the pole toward higher energies with a larger width, which corresponds to a smaller (and potentially unreasonable) $q_{\rm max}$ value. This is why we adopt a small but still physically reasonable value of $0.5$~GeV. It is interesting to note that the bump of the $a_1(1260)$ in the experimental analysis also appears at a much lower mass than 1260~MeV, with a maximum around 1100~MeV, which is consistent with our results. The $\rho^+$ meson, produced in the $a_1(1260)^+$ decay, in turn contributes to the peak observed in the $\pi^+\pi^0$ spectrum.

\subsection{Results in scenario Fixed(X)}

\begin{figure}[h!]
\centering
\includegraphics[bb=5 0 800 1100,clip,scale=0.367]{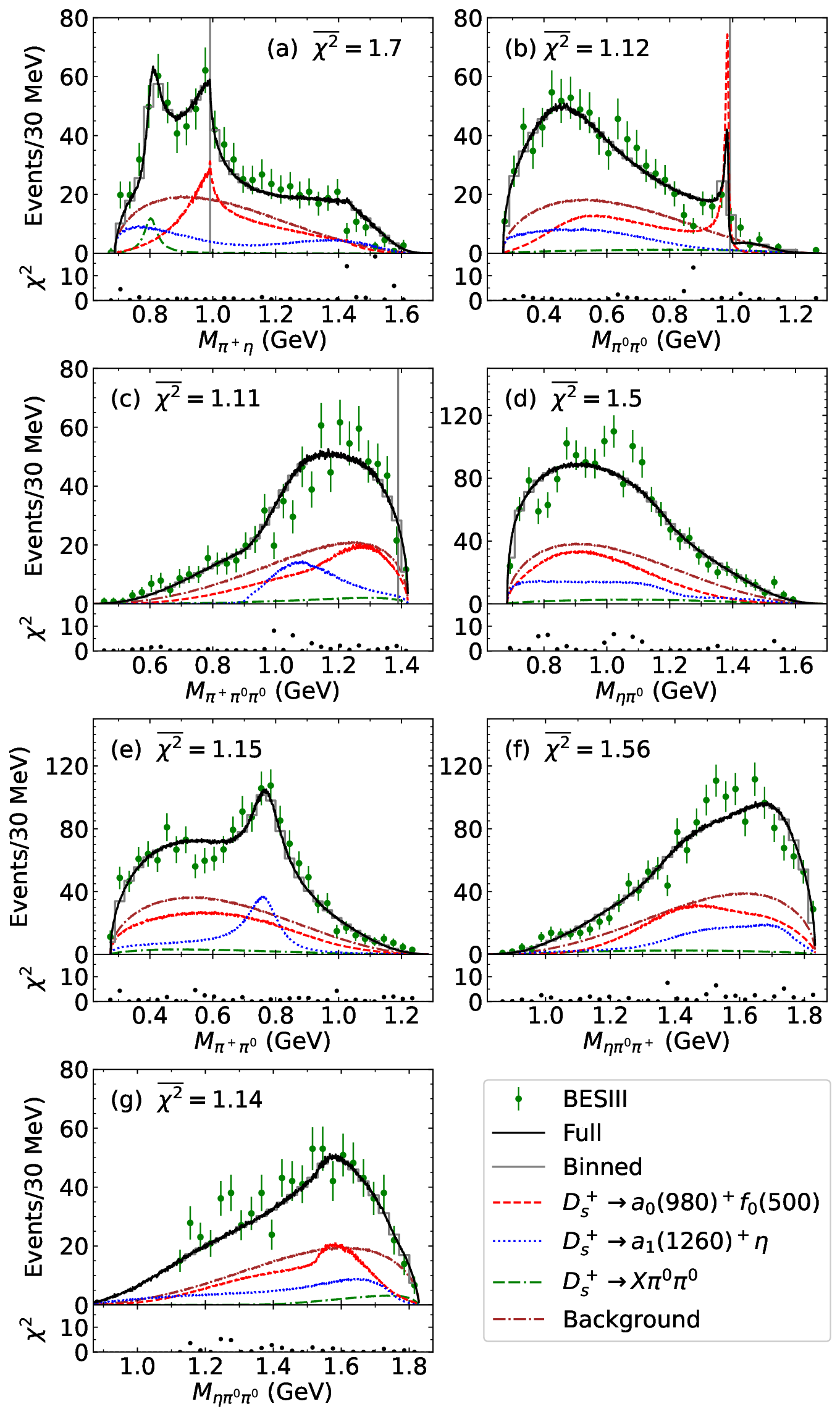}
\caption{The invariant mass spectra of the decay $D_s^+ \to \pi^+ \pi^0 \pi^0\eta$ in scenario Fixed(X). The green dash-dash-dotted line corresponds to the $X(800)$ contribution. Other notations are the same as in Fig.~\ref{fig:Fixed}.}
\label{fig:FixedX}
\end{figure}

We now include the contributions of the $X(800)$ resonance to the $D_s^+ \to \pi^+ \pi^0 \pi^0\eta$ decay. The cutoff parameters $q_{\rm max}$ are still kept fixed at the same values as in the previous subsection, while the strengths $g$, phases $\phi$, and the background parameter are allowed to vary in the fit. The $\chi^2$ per degree of freedom decreases from $1.447$ to $1.398$, indicating only a marginal improvement in the fit quality. The explicit results are illustrated in Fig.~\ref{fig:FixedX}.

With the inclusion of the $X(800)$, a clear peak appears around 800~MeV in the $\pi^+\eta$ spectrum. However, the average $\bar{\chi^2}$ for the $\pi^+\eta$ spectrum only decreases from $1.75$ to $1.70$, reflecting the large experimental uncertainties in the data near 800~MeV. In contrast, more significant improvements are observed in the $\pi^+\pi^0\pi^0$ and $\pi^+\pi^0$ spectra, with decreases of $0.15$ and $0.17$, respectively. This is because the $X(800)$ contribution extends over a broader kinematic region in these spectra, and through interference with other amplitudes, it becomes more influential there. This is precisely why we fix the width of the $X(800)$ to $50$~MeV: to avoid overfitting the data and to ensure that the resonance contributes locally to the $\pi^+\eta$ peak rather than being used to absorb structures elsewhere. Overall, the inclusion of the $X(800)$ does not lead to a significant improvement in the fit, suggesting that, within the fixed-$q_{\rm max}$ model, the current data do not require this additional resonance.

\subsection{Results in scenario Free}

\begin{figure}[h!]
\centering
\includegraphics[bb=5 0 800 1100,clip,scale=0.367]{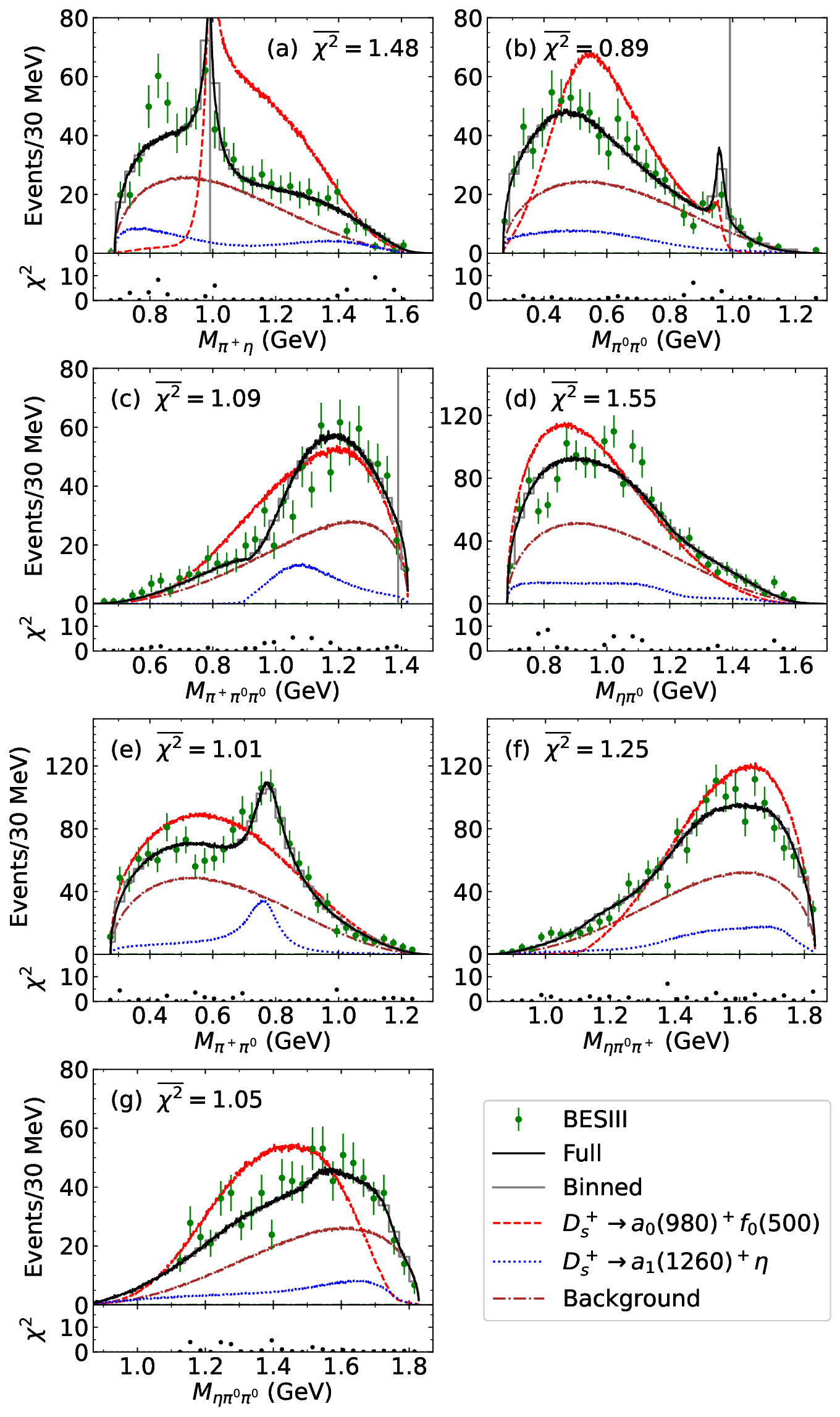}
\caption{The invariant mass spectra of the decay $D_s^+ \to \pi^+ \pi^0 \pi^0\eta$ in scenario Free. Notations are the same as in Fig.~\ref{fig:Fixed}.}
\label{fig:Free}
\end{figure}

In the previous fits, the cutoff parameters $q_{\rm max}$ for the $a_0$ and $f_0$ systems were fixed at 0.6 and 0.65~GeV, respectively. We now release them as free parameters to explore whether a better description of the BESIII data can be achieved. In this subsection, we first omit the $X(800)$ resonance to isolate the effect of varying the cutoffs. The resulting $\chi^2/{\rm ndf}$ is 1.244, which is significantly smaller than the value obtained with fixed $q_{\rm max}$, indicating a notable improvement in the fit quality. The explicit results are illustrated in Fig.~\ref{fig:Free}.

We now discuss the improvements in detail. The peak in the $\pi^+\eta$ spectrum still originates as a cusp effect from the coupled-channel interaction between $\pi^+\eta$ and $K^+\bar{K}^0$. The pole in the second Riemann sheet of the $\pi^+\eta$ channel and the first Riemann sheet of the $K^+\bar{K}^0$ channel is now located at $(1047 - 55i)$~MeV, again arising from the coupled-channel potential. With the best-fitted value $q_{\rm max} = 0.76$~GeV, the peak becomes considerably sharper than that obtained with $0.6$~GeV. This is because a larger cutoff enhances the coupling strength of the coupled-channel dynamics, leading to a more pronounced cusp structure. Unlike the fixed-$q_{\rm max}$ case, the lower side of the peak drops very steeply, while the higher side falls off more gradually.  As a result, the fit becomes more dependent on the interference between different contributions. With this optimized $q_{\rm max}$, the fit to the $\pi^+\eta$ spectrum improves considerably, with the average $\bar{\chi^2}$ decreasing from $1.75$ to $1.48$.

Turning to the $\pi^0\pi^0$ spectrum, with the best-fitted value $q_{\rm max} = 0.854$~GeV for the $f_0$ system, the corresponding pole is located at $(957 - 16i)$~MeV, which is farther from the $K\bar{K}$ threshold and somewhat broader than in the case of $0.65$~GeV. As a result, the peak near 1000~MeV becomes lower and wider, which is in better agreement with the data. The pole corresponding to the $f_0(500)$ moves to $(468 - 211i)$~MeV, providing a broad enhancement in the lower energy region. Since the peak near 1000~MeV is reduced, the low-energy enhancement becomes more pronounced and better matches the data. With this $q_{\rm max}$, the fit to the $\pi^0\pi^0$ spectrum improves markedly, with the average $\bar{\chi}^2$ dropping from $1.16$ to $0.89$.

In addition, the $\pi^+\pi^0\pi^0$, $\pi^+\pi^0$, and $\eta\pi^0\pi^0$ spectra also show noticeable improvements, although the changes are less dramatic than in the $\pi^+\eta$ and $\pi^0\pi^0$ cases. Overall, freeing the cutoff parameters substantially improves the description of the data, suggesting that the fitted $q_{\rm max}$ values are better suited to the dynamics of the $D_s^+$ decay.

\subsection{Results in scenario Free(X)}

\begin{figure}[h!]
\centering
\includegraphics[bb=5 0 800 1100,clip,scale=0.367]{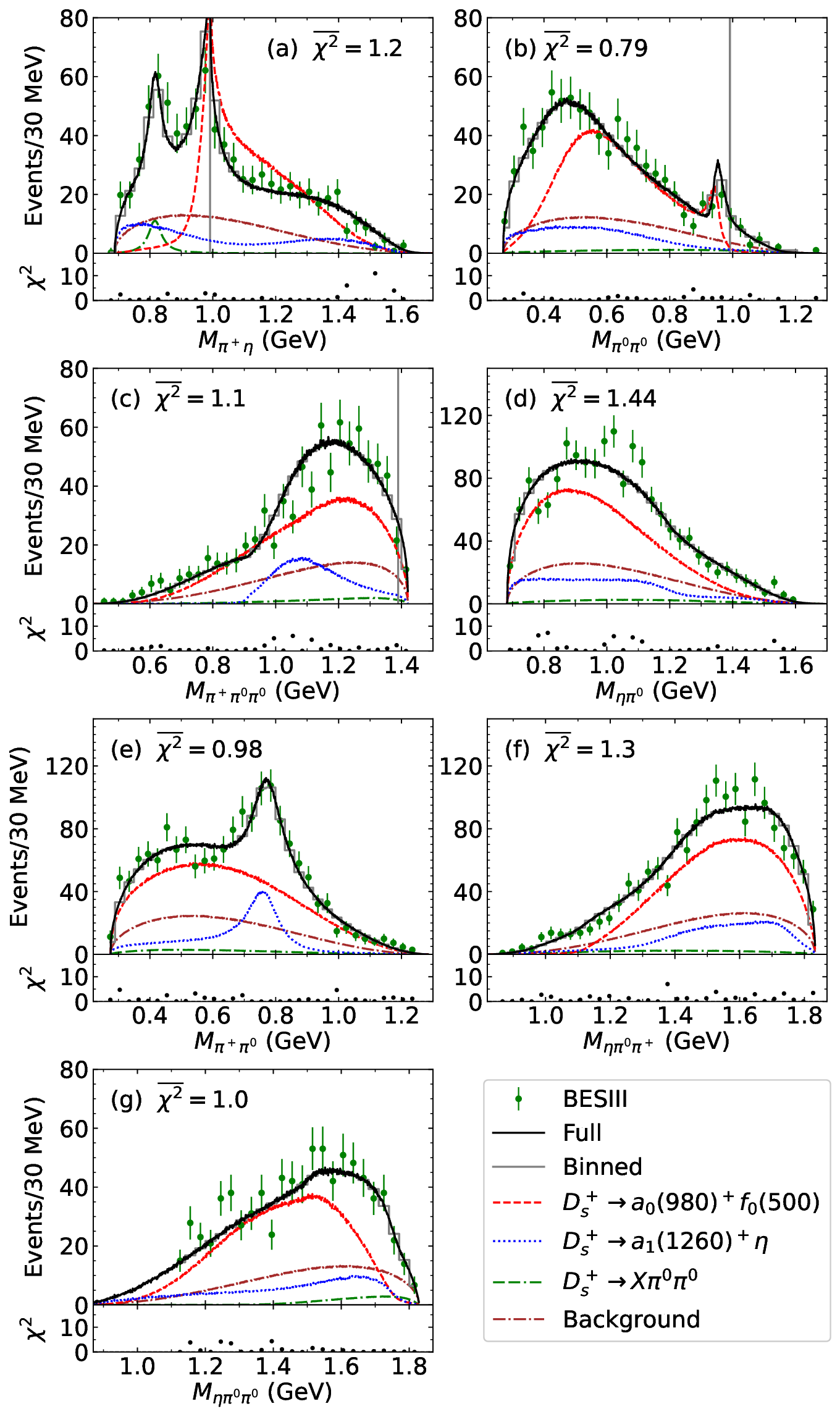}
\caption{The invariant mass spectra of the decay $D_s^+ \to \pi^+ \pi^0 \pi^0\eta$ in scenario Free(X).  Notations are the same as in Fig.~\ref{fig:FixedX}.}
\label{fig:FreeX}
\end{figure}

We now further include the contributions of the $X(800)$ resonance in the model with free $q_{\rm max}$. The cutoff parameters for the $a_0$ and $f_0$ systems remain free, while the $q_{\rm max}$ for the $a_1$ system is kept fixed at $0.5$~GeV as before. The $\chi^2$ per degree of freedom decreases from $1.244$ to $1.190$, indicating only a marginal improvement over the scenario without $X(800)$. The explicit results are illustrated in Fig.~\ref{fig:FreeX}.

With the inclusion of the $X(800)$, a clear peak appears around 800~MeV in the $\pi^+\eta$ spectrum. The average $\bar{\chi}^2$ for this spectrum decreases from $1.48$ to $1.20$, representing a more substantial improvement than in the Fixed(X) scenario, where the corresponding decrease was only from $1.75$ to $1.70$. This difference can be understood as follows: in the Free scenario, with $q_{\rm max} = 0.76$~GeV for the $a_0$ system, the cusp from the $a_0(980)^+$ drops very steeply on the lower side, leaving a region around 800~MeV that is relatively free of contributions from the $a_0$ dynamics. This opens up room for the $X(800)$ to contribute without strongly interfering with the $a_0$ peak. At the same time, the background contribution is reduced accordingly in the fit, allowing the $X(800)$ to account for the excess events near 800~MeV. As a result, unlike in the Fixed(X) case where the improvement was more pronounced in the $\pi^+\pi^0\pi^0$ and $\pi^+\pi^0$ spectra, the most significant improvement here occurs in the $\pi^+\eta$ spectrum itself.

Nevertheless, the overall $\chi^2$ improvement in scenario Free(X) remains marginal. This suggests that even when the cutoff parameters are optimized, the data, given their large uncertainties in the relevant region, do not provide strong evidence for an additional resonance near 800~MeV. We therefore regard the $X(800)$ as a phenomenological addition that marginally improves the local description of the $\pi^+\eta$ spectrum, but is not essential for the global fit.

\section{Summary and discussion}\label{summary}

In this work, we have theoretically investigated the decay mechanism of $D_s^+ \to \pi^+ \pi^0 \pi^0\eta$ based on the recent BESIII data. The decay is described by two distinct dynamical mechanisms: the production of two dynamically generated structures, $D_s^+ \to a_0(980)^+ f_0(500)$, and the production of a single dynamically generated structure, $D_s^+ \to a_1(1260)^+\eta$, followed by the cascade decay $a_1(1260)^+ \to \rho^+ \pi^0 \to \pi^+ \pi^0 \pi^0$. Within the chiral unitary approach, the coupled-channel scattering amplitudes for the $a_0$, $f_0$, and $a_1$ systems, are incorporated them into the decay amplitudes via final-state rescattering. The relative strengths of the various channels are constrained by SU(3) flavor symmetry, while the overall couplings are treated as free parameters fitted to the data.

The fits have been performed on the BESIII invariant mass spectra of $\pi^+\eta$, $\pi^0\pi^0$, $\pi^+\pi^0\pi^0$, $\eta\pi^0$, $\pi^+\pi^0$, $\eta\pi^0\pi^+$, and $\eta\pi^0\pi^0$. From fits under four different scenarios (fixed/free $q_{\rm max}$, with/without $X(800)$), our main findings are as follows:

\begin{enumerate}
\item The peak $a_0(980)^+$ near 1000~MeV in the $\pi^+\eta$ spectrum is well reproduced as a cusp effect arising from the coupled-channel interaction between $\pi^+\eta$ and $K^+\bar{K}^0$, rather than a genuine resonance pole.

\item The structure near 1000~MeV in the $\pi^0\pi^0$ spectrum originates from the dynamically generated $f_0(980)$ state. The broad enhancement at lower energies comes from the $f_0(500)$ pole.

\item The $a_1(1260)^+$ is generated from the  $(\pi\rho)^+$ and $(\bar{K}^*{K} - K^*\bar{K})^+$  interactions. With $q_{\rm max}=0.5$~GeV, the pole is located at $(1010-175i)$~MeV, producing a broad bump around 1050~MeV in the $\pi^+\pi^0\pi^0$ spectrum. The subsequent $\rho^+$ decay accounts for the peak in the $\pi^+\pi^0$ spectrum.

\item Releasing $q_{\rm max}$ for the $a_0$ and $f_0$ systems improves the fit significantly ($\chi^2/{\rm ndf}$ from $1.447$ to $1.243$). In this scenario, interference effects become more important, but the main conclusions remain consistent with those obtained in the fixed-$q_{\rm max}$ scenario.

\item The inclusion of the $X(800)$ resonance, parametrized as a Breit–Wigner, produces a clear peak near 800~MeV in the $\pi^+\eta$ spectrum. However, the overall improvement in the fit is marginal. The limited improvement is mainly due to the large experimental uncertainties in the 800~MeV region of the $\pi^+\eta$ data.
\end{enumerate}

In conclusion, the BESIII data on $D_s^+ \to \pi^+ \pi^0 \pi^0 \eta$ can be well described by the chiral unitary approach, based on two rescattering mechanisms: one involving two intermediate states (the $a_0(980)^+$ cusp and the $f_0(980,500)$ resonance), and the other involving a single intermediate resonance (the $a_1(1260)^+$). The $X(800)$ resonance, although phenomenologically included, does not lead to a substantial improvement. Future high-precision measurements, especially in the low-energy region of the $\pi^+\eta$ spectrum, would be helpful to further clarify the nature of these structures.

\vskip 10pt

\noindent {\bf Acknowledgement} This work is supported by the National Science
Foundation of China (Grant No. 12475080).

\noindent {\bf Data Availability Statement} This manuscript has no associated data or the data will not be deposited. [Authors' comment: This is a theoretical study and no external data are associated with this work.]

\end{document}